# A Polynomial Time Algorithm For Solving Clique Problems (And Subsequently, P=NP)

Michael LaPlante, March 9th 2015

Introduction

Clique problems, such as determining in a given undirected graph of vertices and edges if there is a complete subgraph, or clique, of size k or determining the list of all maximal cliques, have long plagued the computer science field because no known algorithm for solving definitively in an efficient amount of time could be found.

In this paper, I submit an algorithm that efficiently solves the problems k-clique, maximum clique, and complete enumeration of all maximal cliques for any given graph that I have written in the Java programming language. I will first give a high level overview of how the cliques are calculated using this algorithm and show a more in depth look as to how my algorithm works as well as present the code I have written. I will detail how to use it and how to obtain a copy of it.

## 1. Understanding the problem

To understand how the algorithm works, first let us go over in more detail the problem we are trying to solve. A graph can be visualized by a collection of nodes each connecting to other nodes. A clique is a collection of nodes in the graph that each connect to each other.

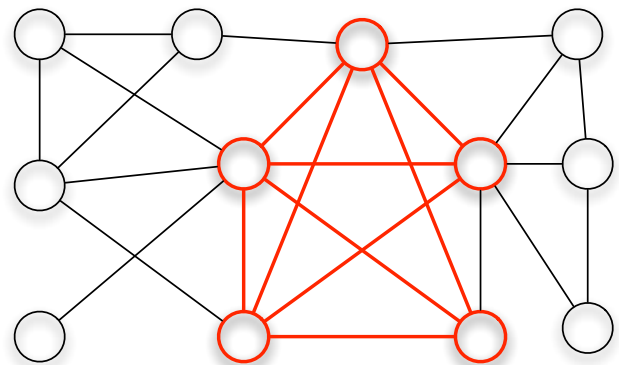

**Figure 1A**

In Figure 1A, each circle is a node in the graph. And each line connecting the node is an edge, representing that the two nodes are connected. The nodes and edges highlighted in red are a part of a clique of 5 nodes. Though there are other cliques in the graph, 5 is the largest clique.

Posing the questions, "What is the largest clique in any given graph?" or "For a given graph, does it contain a clique of size k?" has previously only been known to be solved by searching; that is, iterating through all possible combinations and testing to see if they are a clique. The problem with searching in this fashion is its inefficiency. This takes exponential time based on the size of the graph. In a graph of even 100 nodes, the time to take to calculate cliques in such a graph would take far, far more time than is practical. However, given a list of nodes, we can easily check in polynomial time if they form a clique.

Such is the issue with the class of problems in the category, NP. We are able to check an answer in polynomial time, but unable to find an answer as fast as we can check them.

## 2. How the algorithm works

The approach taken with this algorithm is one with the idea that the nodes can all "communicate" with each other and, almost collectively, decide what cliques they are all a part of. By this communication, each node can first figure out the common 3-cliques (defined as a clique of size 3) that they share with all their neighbor nodes and from there, find out the larger cliques they are a part of.

The heart of the algorithm is in identifying the 3-cliques each node is connected to based on learned mutual neighbors. By each node coming to an understanding of the 3-cliques they are a part of, they can merge these 3-cliques together to calculate the larger cliques. After all, each clique of size 3 or larger is made up of 3-cliques.

This illustration shows a clique of size 4, and the 3-cliques that compose it.

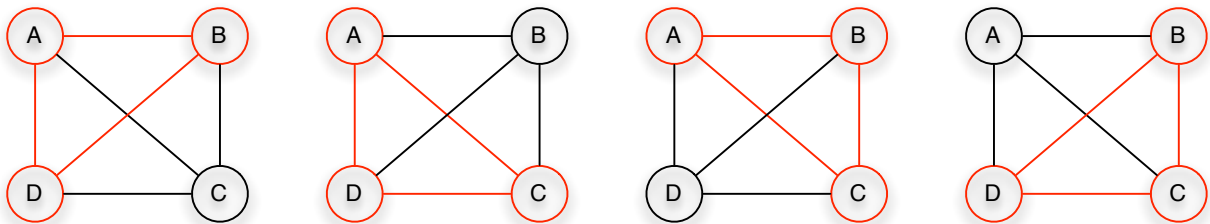

**Figure 2A**

If node A were to learn that it were a part of a 3-clique with nodes B and D, nodes D and C and nodes C and B, then node A would be able to conclude that it must be a part of a 4-clique with them. From Node A's perspective, the mystery of whether nodes B,C,D all know each other is solved because Node A can see from the 3 cliques he calculates, [A,B,D], [A,C,D] and [A,B,C] that the 3-clique, [B,C,D], shows up in them - B and D know each other, C and D know each other and B and C know each other. This exact process will be detailed later.

Ordering of nodes in a clique is irrelevant. Cliques [A,B,C] and [B,A,C] are both the same cliques to this algorithm.

The algorithm works in two phases.

Phase 1:
- Each node tells each of its neighbor nodes, what its own neighbors are.
- Each node upon hearing of a node from a neighbor that it itself has a neighbor with, immediately knows that it is in a 3-clique with those two nodes. This list of discovered 3-cliques is maintained.
- Any nodes that it hears about from these advertisements that it does not have an adjacency with, is disregarded. (although this information could be useful in analysis of disconnected cliques, but that is out of scope for the topic at hand)
- Should a node disregard all of the nodes advertised from a neighbor node, the node knows it is a part of a maximal 2-clique with that node.

Phase 2:
- After completing phase 1, each node now has a table of the 3-cliques it is a part of.
- Using this information, each node can merge the 3-cliques into larger cliques.
- This merging is accomplished by selecting key nodes and determining which nodes share a connection with the key node and merging those nodes into a list containing the key node if there exists a pair for each node in the list and the new node. *(this may sound complicated now, but the method proposed and detailed later is quite simple)*

## 3. Detailed Explanation of the Algorithm

Phase 1 - Building the knowledge of a given node's 3-cliques to which it belongs:

This first phase makes a pass on each of the nodes in the graph. For each of these nodes, the adjacent nodes are advertised to the each of the other neighbors. Each node, learning about the neighbors of its own neighbors, can determine the list of 3-cliques it belongs to based on if it also is a neighbor of the advertised neighbor.

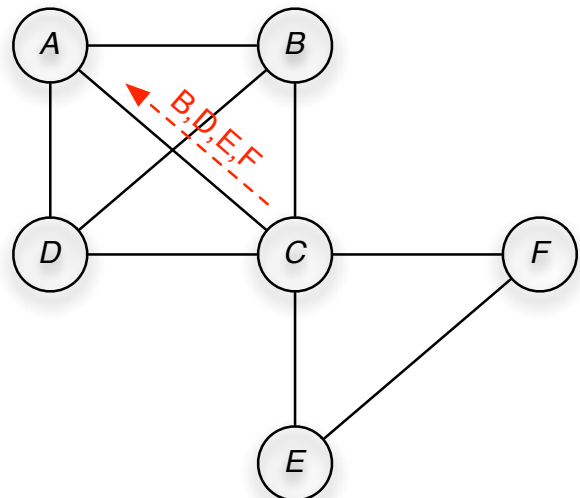

**Figure 3A - Node C advertises its adjacent nodes B,D,E and F.**

For example, in figure 3A, Node C advertises to Node A, its directly connected neighbors B,D,E,F. Node A sees that B is a neighbor of its own, therefore it concludes that there is a 3-clique with B and C and adds to its list of 3-cliques. The same thing happens when it sees D. Upon learning that C has neighbor E and that A does not have E as a neighbor, A disregards this information. The same thing happens with F. Should a node learn there are no mutual neighbors with a particular neighbor, the node can conclude it is a part of a maximal 2-clique with that neighbor node.

Figure 3B shows Node A's list of 3-cliques it knows it belongs to and the list of mutual neighbors with node C after this part of the process.

Figure 3C shows the end calculation of each node from figure 3A. Note that the 3-cliques listed for each node, is the entire list of 3-cliques it belongs to after hearing of the neighbor advertisements from each of its neighboring nodes.

This is the extent of the logic involved in the first phase. Each node now has a good picture of their "neighborhood" and which neighbors know each other. I call this process "Neighbor Introductions." By following this process we have effectively isolated the appropriate subsets of the entire graph, eliminating the requirement of searching and testing completely unrelated nodes or following paths of connected nodes searching for whether neighboring nodes know each other. Building this data of the neighborhood is a simple process, easily calculable in polynomial time. The logic in the 2nd phase, which I call "Clique Calculation," is relatively simple as well.

| | Mutual Neighbors with C | B,D |
|---|---|---|
| | 3-cliques calculated from C's advertisements | [A,B,C], [A,C,D] |

**Figure 3B - Node A's calculated 3-cliques**

| Node | 3-cliques | Neighbor | Mutual Neighbors |
|---|---|---|---|
| A | [A,B,C], [A,B,D], [A,C,D] | B | C,D |
| | | C | B,D |
| | | D | B,C |
| B | [A,B,C], [A,B,D], [B,C,D] | A | C,D |
| | | C | A,D |
| | | D | A,C |
| C | [A,B,C], [A,C,D], [B,C,D], [C,E,F] | A | B,D |
| | | B | A,D |
| | | D | A,B |
| | | E | F |
| | | F | E |
| D | [A,B,D], [A,C,D], [B,C,D] | A | B,C |
| | | B | A,C |
| | | C | A,B |
| E | [C,E,F] | C | F |
| | | F | C |
| F | [C,E,F] | C | E |
| | | E | F |

**Figure 3C - After each node has heard from each of its neighbors, a complete view of the 3-cliques it belongs to has been built.**

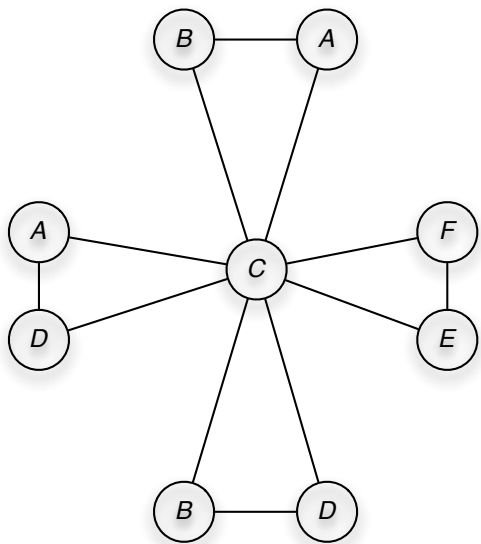

**Figure 3D - C's View of its 3-cliques**

Figure 3D shows how node C see's the 3-cliques he is a part of. In the next phase, each node utilizes their own compiled data to merge the 3-cliques together to calculate the larger cliques. Each node will iterate over each pair of nodes in the 3-clique list to figure out where they fit in the larger cliques it may be a part of. The pairs that C sees in this example are [A,B], [A,D], [B,D] and [F,E]. For this example, we'll do this from node C's perspective, but for the entire algorithm to work, of course, each node will have to the same steps.

First we'll take the first pair we'll analyze. The order does not matter. Arbitrarily we'll choose the top pair, B and A. From that pair we also identify a key node - the node we'll do the merging on. A key node is chosen, because if that node is a part of a larger clique, then those other nodes will also have a pair with that key node as per the definition of a clique.

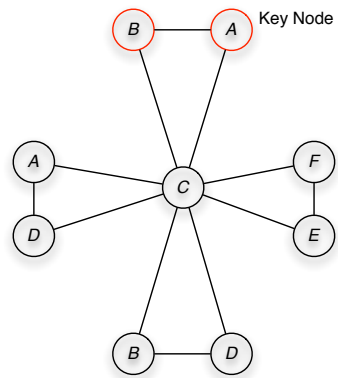

**Figure 3E - Pair B,A is chosen**

Next, we iterate the list of other pairs, searching for a pair that has our key node. That would be pair A,D.

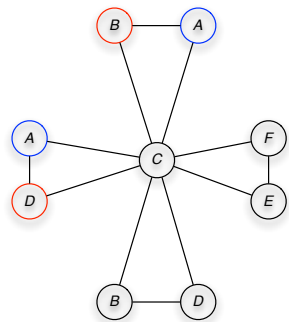

**Figure 3F - A,D has been identified since it is the next pair which has the key node A in it.**

Before we can conclude that pair A,D can be merged we have to determine if the other node in this pair (node D) shares a pair with each node in the list of the other nodes, currently node B in this state. This is the case, because we see the pair B,D - see figure 3G.

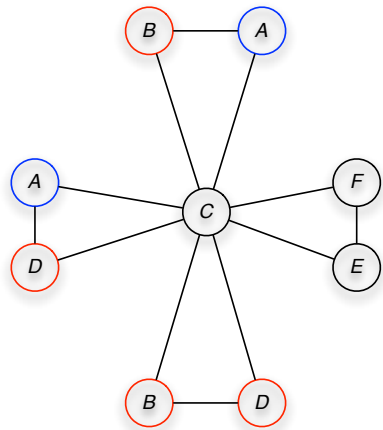

**Figure 3G - D is in a pair with B**

We now can merge these found pairs into our clique as shown in figure 3H effectively adding D to the list with C,A,B.

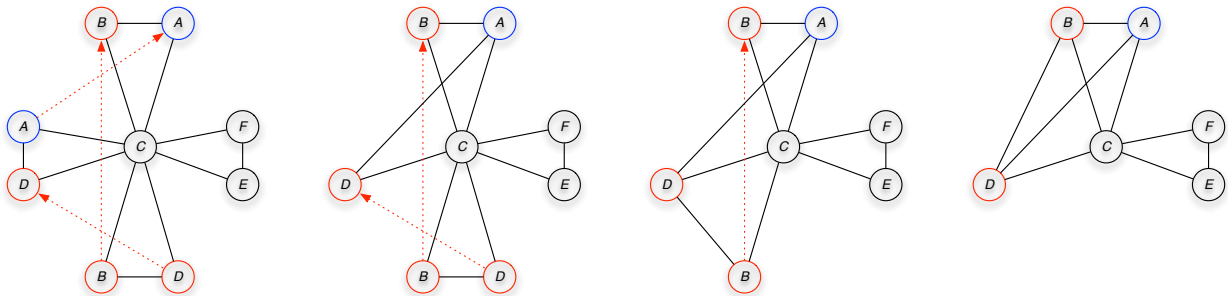

**Figure 3H - The found pairs are merged.**

We have concluded that C is at least a part of a 4-clique containing the nodes A,B,D and itself. We shall continue the process to find out if the other remaining pairs can be merged into this 4 clique as well. In this case, no remaining node pair, (F,E is the last one) has the key node A. Therefore we are finished with one clique. Because there are no other nodes to merge into this clique, [C,A,B,D] is a maximal clique. We can now continue to find out the other cliques that can be merged.

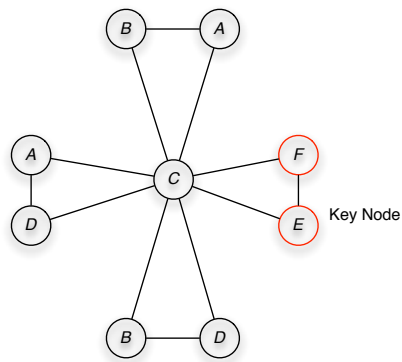

**Figure 3I - Pair [E,F] is the next pair to be analyzed with Node E as the key node**

The next node pair to analyze will be one that was not merged into a previous clique. In this example, pair [E,F] is the next pair. We select E as the key node.

Because no other pair has the key node, we can conclude that [C,E,F] cannot be merged with anything else and [C,E,F] is a maximal clique. Note that the next node pair to be analyzed was not a node pair that was merged into a larger clique, however the next node pair to be analyzed still has access to the other node pairs to determine which ones can be merged onto the key node. That part of the logic will become more apparent in the next section.

I feel the beauty of this algorithm is that it doesn't matter what the graph as a whole looks like. For any given node, all the data that needs to be compiled is the directly adjacent neighboring nodes, and the directly adjacent neighboring nodes of those. In the next section, I will show a more complex scenario to illustrate this further.

## 4. A more complex example to show the resiliency of the algorithm

Consider the graph in Figure 4A. We will view the calculations from node C's perspective, but of course, each node will perform their own calculations using the same process to derive their own independent results.

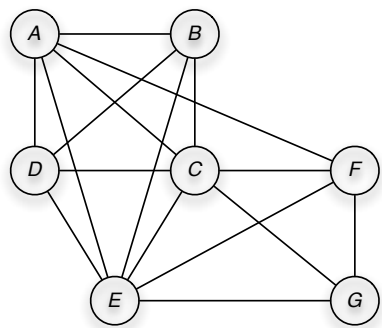

**Figure 4A**

This graph is a little more complex in that node C is a part of 5-clique (A,B,C,D,E) and 4-clique (C,E,F,G) and 4-clique (A,C,E,F) with node E showing up in each clique and nodes F and A showing up in 2 of them. Using the same algorithm, we'll show that the correct results will be determined.

Phase 1 - Neighbor Introductions

Node A tells Node C about its neighbors B,D,E,F. Node C, having a connection to each of those neighbors, determines it has the following 3-cliques. [C,A,B], [C,A,D], [C,A,E], [C,A,F].

Node B tells Node C about its neighbors A,D,E. Node C, having a connection to each of those neighbors, determines it has the following additional 3-cliques. [C,B,D], [C,B,E]. The 3-clique [C,B,A] is not added because it is the equivalent of [C,A,B], which it already has.

Node D tells Node C about its neighbors A,B,E. Node C, having a connection to each of those neighbors, determines it has the following additional 3-cliques. [C,D,E]. Once again, the 3-clique, [C,D,A] is the equivalent of [C,A,D] and [C,D,B] is the equivalent to [C,B,D], both of which were learned earlier.

Node E tells Node C about its neighbors A,B,D,F,G. Node C, having a connection to each of those neighbors, determines it has the following additional 3-cliques, [C,E,F], [C,E,G]. Yet still, 3-cliques ([C,E,A], [C,E,B], [C,E,D]) are equivalent to existing 3-cliques, so they are not added.

Node F tells Node C about its neighbors A,E,G. Node C, having a connection to each of those neighbors, determines it has the following additional 3-cliques. [C,F,G] once again, because the other 3-cliques are equivalent to existing 3-cliques, only that 3-clique is added.

Node G tells Node C about its neighbors E,F. Node C concludes no additional 3-cliques, once again, because 3-cliques [C,G,E] and [C,G,F] are equivalent to already learned ones.

This entire list of 3-cliques is, [C,A,B], [C,A,D], [C,A,E], [C,A,F], [C,B,D], [C,B,E], [C,D,E], [C,E,F], [C,E,G], [C,F,G]

Figure 4B is now Node C's view of its "neighborhood" and Phase 2 can begin.

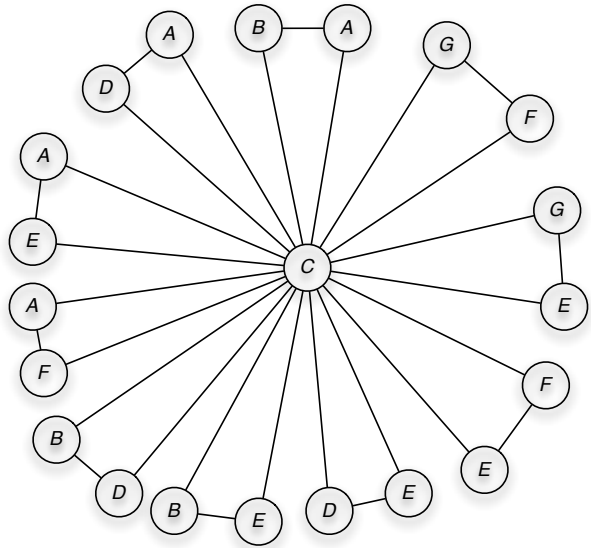

**Figure 4B**

Phase 2 - Clique Calculation

Starting arbitrarily with the top pair B,A we identify Node A, also arbitrarily, as the key node. See figure 4C.

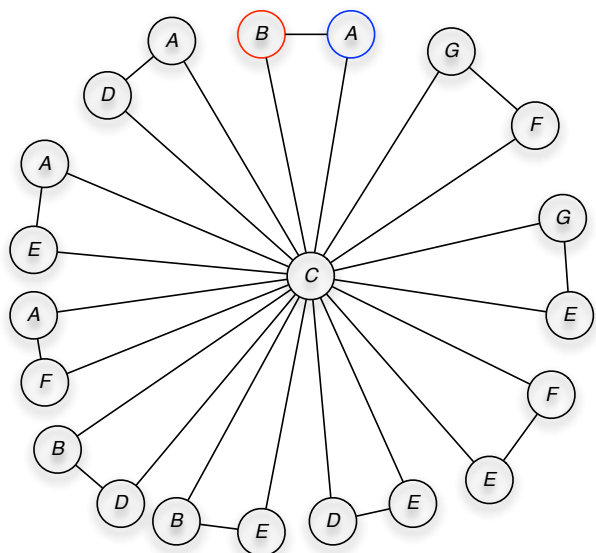

**Figure 4C**

The first pair to check for if a key node exists is A,D and it does indeed have the key node, A.

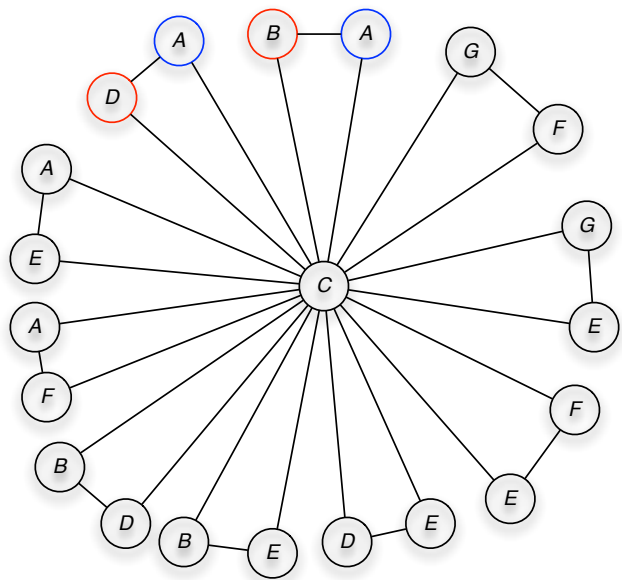

**Figure 4D**

Of course before concluding we can add D to the clique with A,B we have to find a pair with D,B. Which we do.

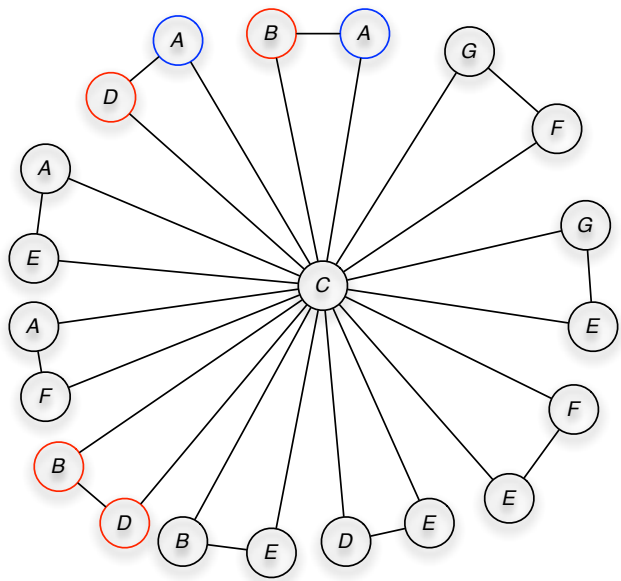

**Figure 4E**

So A,B, A,D and B,D are merged.

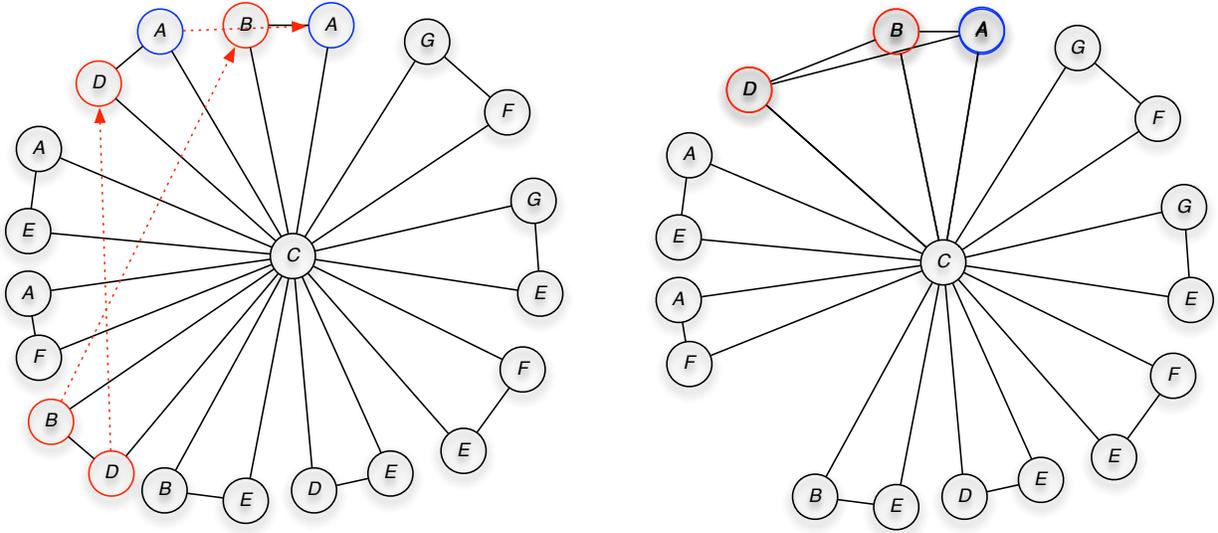

**Figure 4F**

The next pair processed is A,E, which has key node A in it. But before deciding to add E to the list A,B,D we must ensure there is a pair for every other node in the list, B,D.

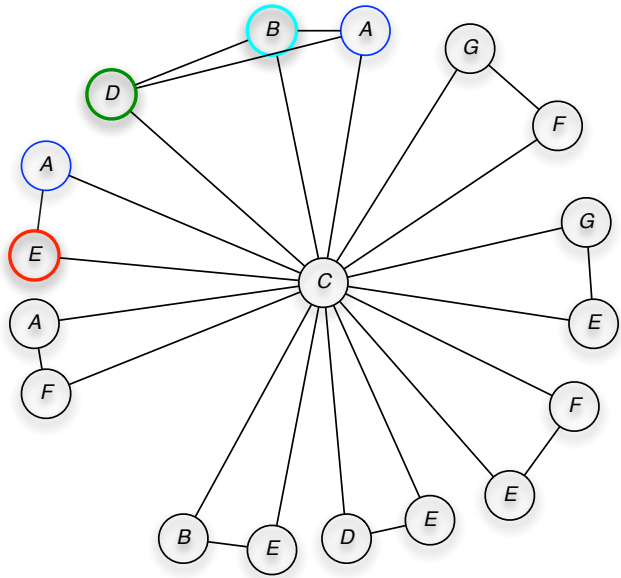

**Figure 4G**

Indeed, node pairs B,E and D,E do exist.

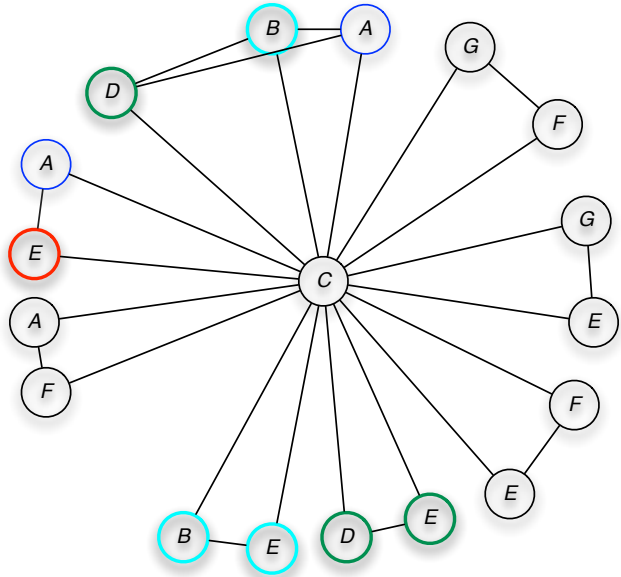

**Figure 4H**

And so these pairs are all merged.

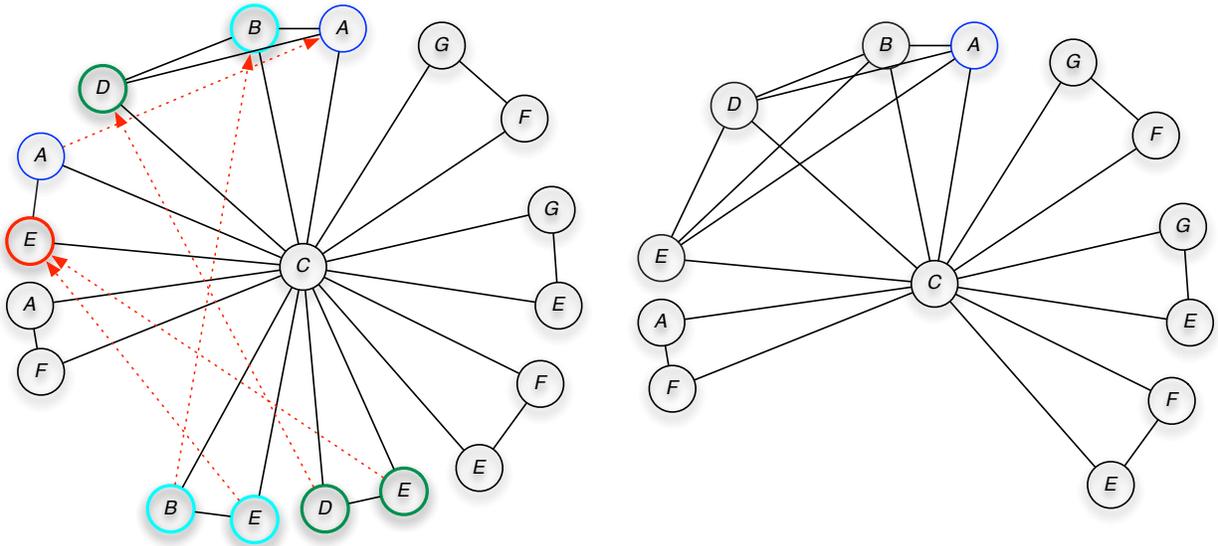

**Figure 4I**

Next we analyze pair A,F because it has the key node in it.

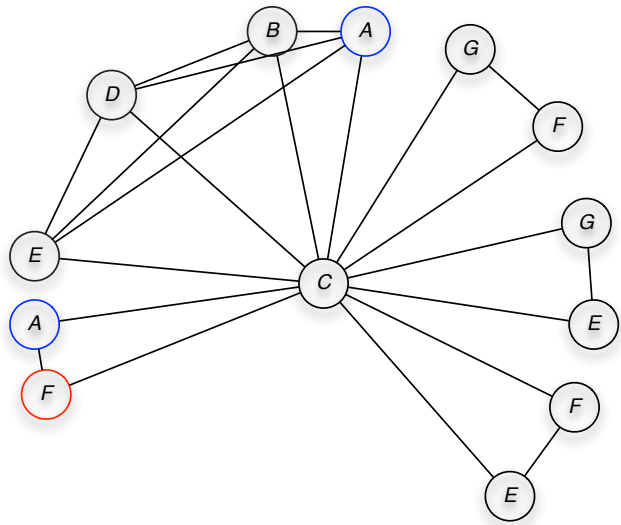

**Figure 4J**

As the algorithm dictates, we must find a pair of nodes for each of the nodes in the merged clique and this new node (F) paired with the key node. Pair E,F exists, but neither D,F nor B,F do. It only takes the lack of one pair to conclude we cannot add F to the clique.

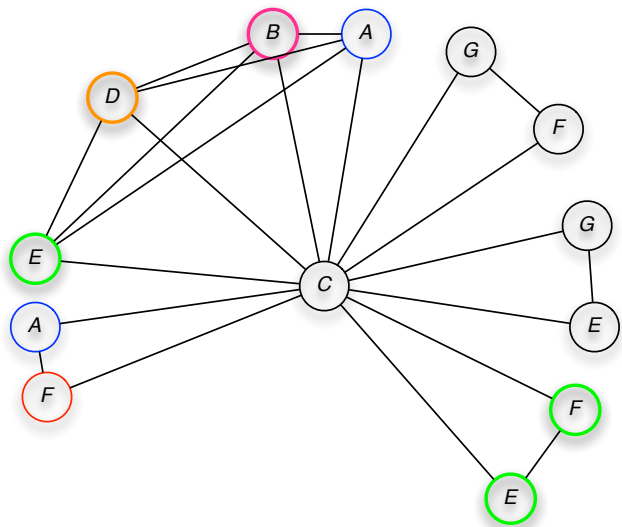

**Figure 4K**

So we move on and attempt to find another pair with the key node in it. No other pair is found with A in it. Therefore this merged clique is complete and consists of nodes [C,A,B,D,E] and is a maximal clique.

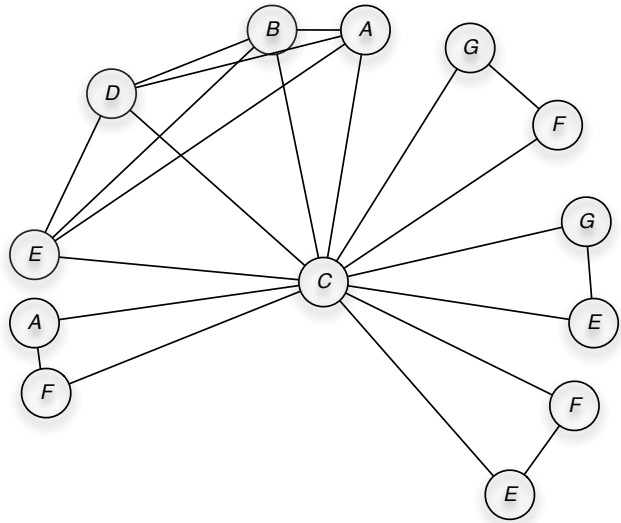

**Figure 4L**

Continuing on, we select the next pair that has not been merged. This is pair A,F. We shall, arbitrarily, designate F as the key node.

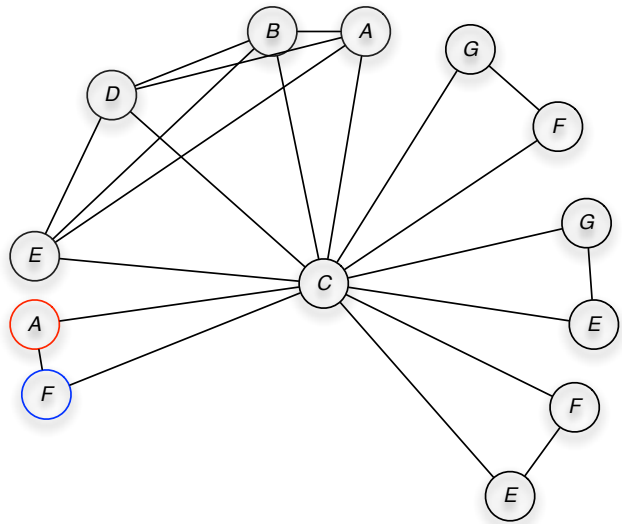

**Figure 4M**

Using F as a key node, we shall look for a pair containing F. We do have access to each pair, even the ones that were merged previously. Pair F,E is found.

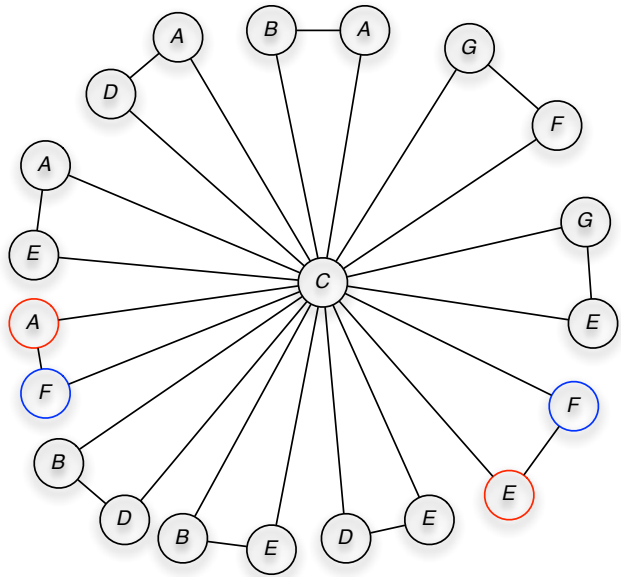

**Figure 4N**

And indeed, there is a pair A,E which is a requirement to add E to the clique with A,F. These nodes are merged into the list as before.

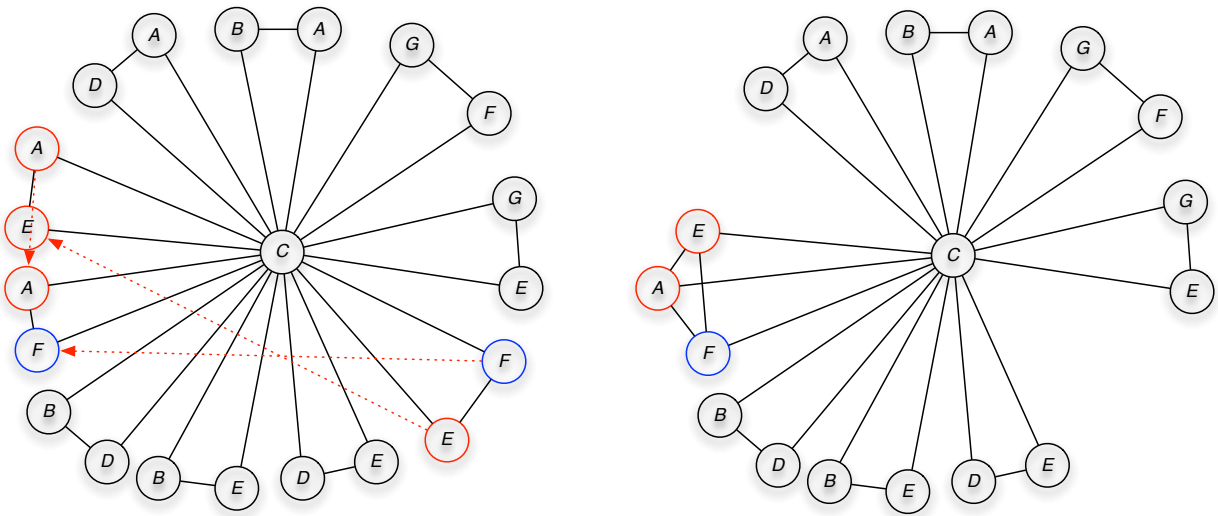

**Figure 4O**

There is another pair with F as the key node, F,G. However, while there does exist a pair G,E, pair G,A does not exist. Again, it only takes the lack of one pair of nodes in the list of merged nodes and the new node to add to determine to not add the node. G is not added. There are no more nodes with F as a member. [C,F,E,A] is a maximal clique.

The next node pair that has not been merged is selected and a key node is identified. The order of the pairs we choose is arbitrary, so let us choose F,G and key node F again to see what happens with this algorithm.

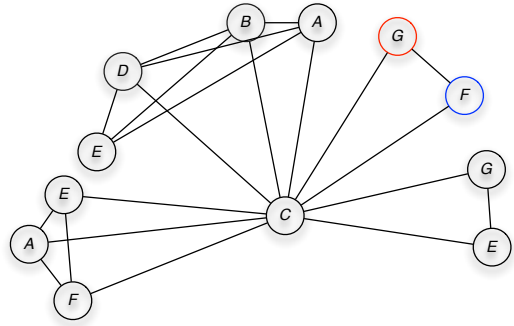

**Figure 4P**

As before, we have access to each of the pairs, even the pairs previously merged. Looking for a pair with the key node F, we find pair F,A. Before merging, we have to see if there exists a pair G,A, which there is none. Therefore pair F,A is skipped.

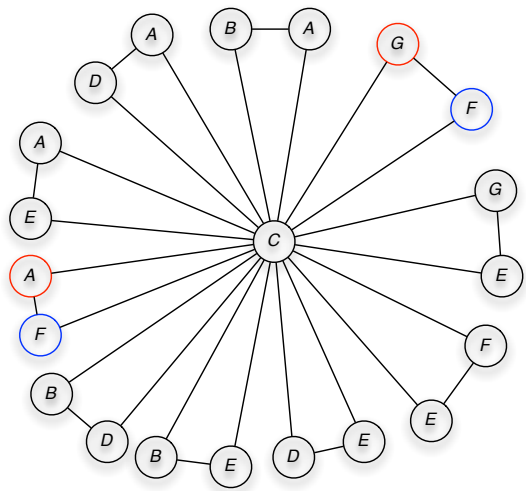

**Figure 4Q**

F,E is the next node pair with the key node in it. There is also a pair G,E, therefore we can merge these pairs.

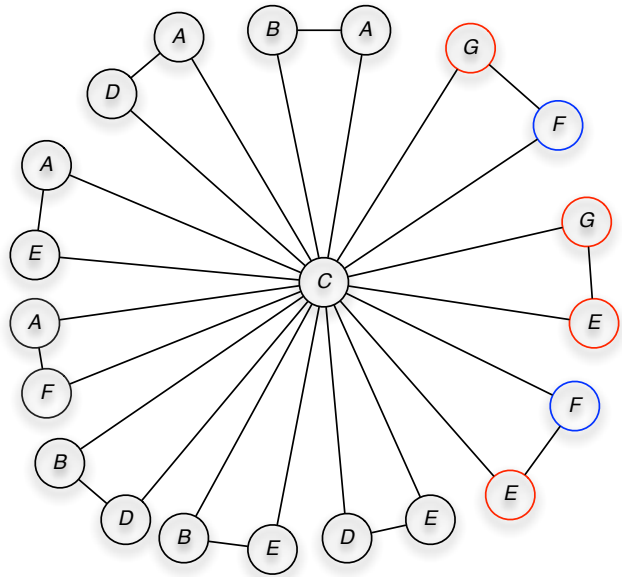

**Figure 4R**

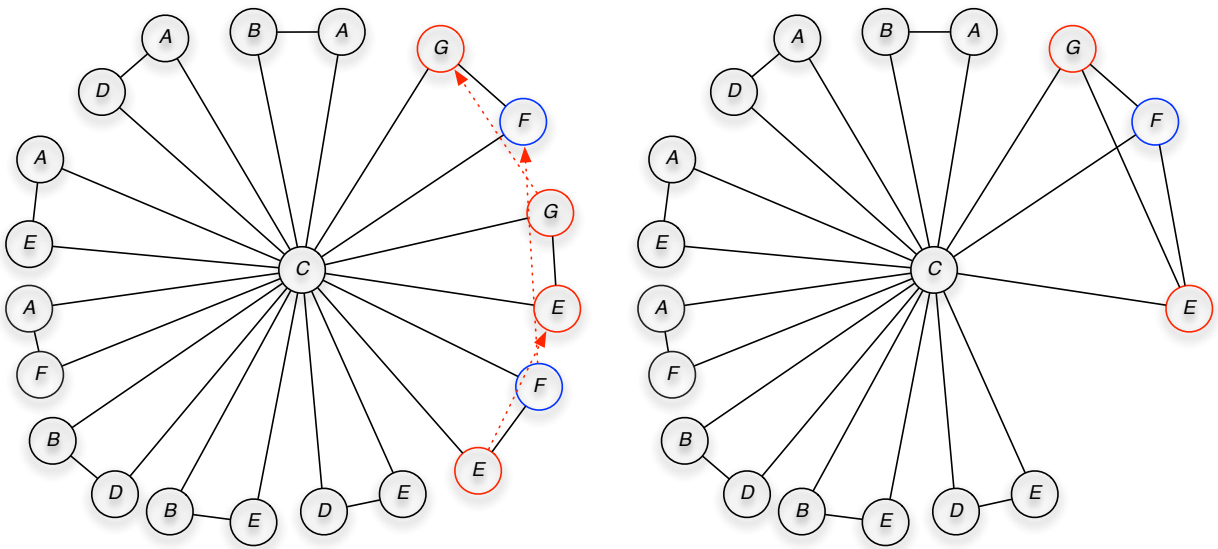

**Figure 4S**

There are no other pairs with key node F in them, so this clique, [C,E,F,G] is maximal. Figure 4T shows the final conclusion C draws about the separate maximal cliques it belongs to.

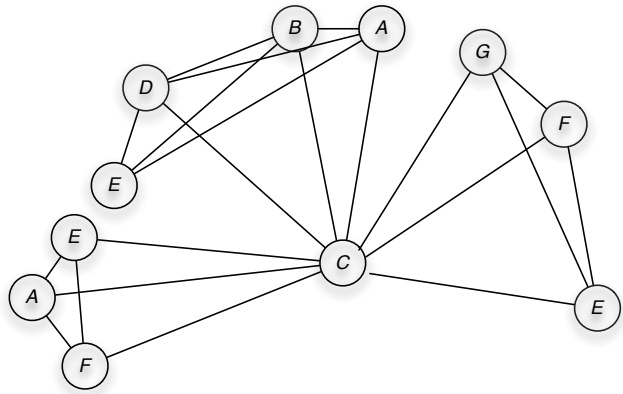

**Figure 4T**

Node C has properly constructed the 3 separate cliques it belongs to, a 5-clique and two 4-cliques. Node A shows up in two of them, node E shows up in all of them and node F shows up in two of them, all as we expected.

As I have mentioned before, I feel the beauty of the algorithm is that it works no matter the type of graph input to it. By the initial compilation of the necessary data, easily done in polynomial time, for each node the algorithm can derive the separate maximal cliques to which they each belong, also easily done in polynomial time. I will show the implementation in pseudo-code and show where a working version written in Java, but first I will show what questions can be answered after each node has performed these calculations.

## 5. Solving the problems "What is the largest clique size?" and "What are all the maximal cliques?" and the decision problem "Does there exist a clique of size k?"

Solving these problems become trivial tasks at this point. The decision problem, "Does there exist a clique of size k?" can be determined (as well as the nodes belonging to that clique can be returned) at the instance a full clique is determined. Each time a node finishes a clique calculation, the algorithm can check to see if the size of that clique is ≥ k. If so, then it can return "YES" and/or the nodes of that calculated clique.

To determine the largest clique size of the given graph, each time a new clique (C) is found, the size of that clique can be determined and compared to previously largest clique (L) found. If the size of C is > the size of L, then C replaces L as the largest one found. After each node has calculated their cliques, the largest size is returned.

Each merged clique returned by the algorithm is a maximal clique. This is because all node pairs that can be merged onto the key node, are merged; none are left out. Should a "non-mergable" 3-clique be found, it is a maximal 3-clique. During phase 1 of the algorithm, where the nodes introduce their neighbors to each of their other neighbors, should a node find it shares no common neighbors with one of its neighbors, it can be concluded that those two nodes are in a maximal 2-clique. While each node will independently calculate the cliques it is a part of, only unique cliques are returned by the main function of the algorithm. For example, cliques [A,B,C,D] and [B,C,A,D] are both considered the same clique by the algorithm and are not unique. Therefore, the answer to the question, "What are all the maximal cliques?" is, every clique the algorithm finds.

## 6. The algorithm in Pseudo-code and an analysis of the algorithm's run time

A complete and working version of this algorithm was developed in Java and posted online. Details on that will be provided later in this paper. This is the pseudo-code of the algorithm.

A crucial part to the algorithm is the list of neighboring nodes that each node has. In my implementation, I have gotten that information "for free" when I build the graph. The algorithm maintains a list of neighbors on each node and adds to it, the new neighbor when connecting 2 nodes together. Observe the below pseudo-code.

```
// An implementation of a graph method that, as part of building out a graph,
// connects two nodes together and tells each node they are now neighbors.
FUNCTION CONNECT_NODES(a,b)
        TELL NODE a IT IS NEIGHBORS WITH NODE b
        TELL NODE b IT IS NEIGHBORS WITH NODE a
        // do other graph stuff, such as keep track of edge count
END FUNCTION
```

But, of course, if one is inclined to do this process separately, it will take `O(n(n-1))` for a complete graph where `n` is the number of nodes in the graph for each neighbor to build its connected neighbor list.

```
// PHASE 1 - Neighbor Introductions
// A Node performing neighbor introductions, that is, telling
// each node that it shares an edge with about all the other
// nodes it shares an edge with
FOR EACH NODE n IN THIS NODE'S NEIGHBORS
        3CLIQUE_COUNT = 0
        FOR EACH NODE nn IN THIS NODE'S NEIGHBORS
                IF (n DOES NOT EQUAL nn) THEN
                        TELL NODE n THAT THIS NODE KNOWS nn
                        IF n KNOWS nn ALSO THEN
                                3CLIQUE_COUNT = 3CLIQUE_COUNT + 1
                        END IF
                END IF
        NEXT NODE
        IF 3CLIQUE_COUNT = 0 THEN
                // we just found out we are a part of a maximal 2-clique with node n
                ADD CLIQUE WITH THIS NODE AND n TO LIST OF MAXIMAL CLIQUES
        END IF
NEXT NODE

// A Node hearing about a neighboring node's neighbor from the
// the above pseudo-code
// n = the neighbor doing the introduction
// nn = the neighbor of n being introduced
FUNCTION INTRODUCED_TO_A_NEIGHBOR'S_NEIGHBOR(NODE n, NODE nn)
        IF THIS NODE HAS nn AS A NEIGHBOR THEN
                IF 3_CLIQUE (THIS NODE, n, nn) DOES NOT EXIST THEN
                        CREATE_NEW_3CLIQUE_WITH_NODES(THIS NODE, n, nn)
                END IF
        END IF
END FUNCTION
```

Each node performs the above operations. The length of time taken to do the above task is based on the number of edges. Each node will do this for each edge, therefore in a complete graph each node has `n-1` edges where n is the number of nodes. Each node will be telling each neighboring node about the other neighbors, of course not itself and the other node, therefore that calculation is done `(n-2)(n-3)` times. Therefore the Big O notation for phase one is at most `O((n-1)(n-2)(n-3))` for a complete graph.

```
// PHASE 2 - Clique Calculations
// A Node processing all of the 3-cliques that were determined
// in phase 1, merging them into the larger cliques they are
// all a part of

// First, build a list of node pairs to evaluate based on the
// 3-cliques calculated in phase 1
FOR EACH 3CLIQUE IN LIST_OF_THIS_NODE'S_3-CLIQUES
        // if this node is node A and there is a 3-clique [A,B,C]
        // we want to add pair [B,C] to the list of pairs
        ADD PAIR(n, nn) WHERE NEITHER n NOR nn ARE NOT THIS NODE TO PAIRLIST
NEXT 3CLIQUE

// Now we perform merging
FOR EACH NODE_PAIR p IN PAIRLIST WHERE p IS NOT MARKED AS MERGED
        CREATE LIST OF NODES CALLED list
        ADD BOTH NODES IN p TO list
        SET KEYNODE TO BE THE FIRST NODE IN p
        FOR EACH NODE_PAIR pp IN PAIRLIST WHERE pp IS NOT p // analyze all other pairs
              // Check to see if the pair has the keynode
              IF KEYNODE IS IN pp THEN
                    // check to see if the other node in pp
                    // is also paired with every other node
                    // in list
                    NODE s = OTHER_NODE_IN pp // s will not be the key node
                    FOR EACH NODE n IN list
                            // before adding s to list, we need to make sure
                            // each node in list has a pair with s
                            DETERMINE_IF_NODES_PAIRED_IN_PAIRLIST(s,n)
                    NEXT NODE
                    IF ALL NODES IN list ARE PAIRED WITH s THEN
                            MARK EACH PAIR PAIRED WITH s AS MERGED
                            MARK PAIR pp AS MERGED
                            ADD s TO list
                    END IF
              END IF
        NEXT NODE
        CREATE NEW_CLIQUE WITH THIS NODE AND EACH NODE IN list
        IF WE ARE SUPPOSED TO BREAK UPON FINDING A CLIQUE OF A GIVEN SIZE THEN
              IF NEW_CLIQUE >= SIZE_CRITERIA THEN
                    END ALGORITHM
              END IF
        END IF
NEXT NODE
```

This process is done on each node pair that shows up in the 3-clique list for each given node. In a complete graph for any given node N, there are `((n-1)(n-2))/2` (where n is the number of nodes in the graph) 3-cliques that N will belong to. For a complete graph, this process will be fastest. It will only take about `O(n((n-1)(n-2))/2)` where n is the number of nodes in the graph. Comparing each pair against each other pair to determine "mergability" takes at most $O(n(p(p-1)))$ where p is the number of pairs that show up in the pair list for a node determined by the 3-cliques and n is the number of nodes in the graph.

Note that the perspective of each node performing all of this logic is arbitrary. It could easily also be viewed as an observer of the graph compiling and doing all of the calculations. I feel it helps to visualize the algorithm as being seen as an entity inside the graph executing the calculations.

## 7. Implementation of this algorithm built in the Java programming language

As mentioned before, I have built a working implementation of this algorithm in Java. That implementation has been uploaded to the free code sharing website, github.com, and can be obtained at the following URL.

http://github.com/mlaplante1981/CliqueProblem

Details on how to use it are at the same location, but an overview of its use is as follows.

You will first need to create a new blank Graph object.

```
Graph graph = new Graph();
```

Creating Nodes in the graph can be done one of several ways.

Create nodes individually.

```
Node a = graph.newNode(); // brand new node in the graph, not connected to anything
Node b = graph.newNode(); // brand new node in the graph, not connected to anything
```

Or by creating a list of nodes of a given size.

```
NodeList list = graph.newNodes(5); // creates 5 nodes, none connected to anything yet
```

The above created list is a zero-based array object. Accessing each node in the above created list is done by:

```
Node c = list.get(0);
Node d = list.get(1);
```

Connecting nodes is done by telling the graph to connect the nodes.

```
graph.connectNodes(a,b); // Nodes a and b are now connected to each other
```

For convenience, there is a method to create a complete sub-graph of a given size.

```
NodeList cl5 = graph.newCompleteGraph(5); // returns a list of nodes, each connecting
                                          // to each other - a 5 clique here
```

Each of these nodes can be connected to any other nodes.

```
graph.connectNodes(cl5.get(1),a,c); // the node indexed at 1, a and b now each connect
                                    // to each other
```

To perform the calculation, call the method doCalculation() on the graph object

```
graph.doCalculation();
```

After calculations are completed, the largest clique, its size and each of the maximal cliques can be obtained.

```
Clique largestClique = graph.getLargestFoundClique(); // returns the largest clique
int size = largestClique.size(); // returns the size of the largest clique
Set<Clique> allMaximalcliques = graph.getCliquesFound();// returns all cliques, each
                                                       // of which is a maximal clique
```

If you want the algorithm to stop executing upon finding a clique of a certain size:

```
graph.setBreakOnCliqueOfSizeFound(50); // algorithm stops if it finds a clique size 50
```

To view the output of each node's calculated maximal cliques, call `printOut()` on graph *after* `doCalculation()`.

```
graph.printOut(); // prints out each of the cliques that each node calculates
```

An example of solving 3SAT reduced to Clique in semi-pseudocode:

```
// Create a node for each variable in the clauses
FOR EACH CLAUSE clause
       SET var1 IN clause TO graph.newNode();
       SET var2 IN clause TO graph.newNode();
       SET var3 IN clause TO graph.newNode();
NEXT CLAUSE
// Connect each of the nodes together in the graph
FOR EACH CLAUSE clause
       FOR EACH OTHER CLAUSE otherClause
         // calling graph.connectNodes() for each of the following connections
         CONNECT clause.var1 to each VAR in otherClause ACCORDING TO REDUCTION RULES
         CONNECT clause.var2 to each VAR in otherClause ACCORDING TO REDUCTION RULES
         CONNECT clause.var3 to each VAR in otherClause ACCORDING TO REDUCTION RULES
       NEXT CLAUSE
NEXT CLAUSE
graph.doCalculation();
if (graph.getLargestClique().size() == clauseCount){// this 3SAT is satisfiable!}
```

## 8. Application to the open question on P vs NP

The question on whether or not an algorithm exists to find an answer in polynomial time to a searching problem if the answer can be verified in polynomial time has been around since the 1970's. Stephen Cook essentially posed the question initially in his paper, "The complexity of theorem proving procedures" in 1971[1]. Leonid Levin also independently laid the ground work for this question in his paper "Universal search problems" in 1973[2]. Richard M. Karp showed how to reduce satisfiability to clique as well as defining 19 other NP-Complete problems in his 1972 paper, "Reducibility Among Combinatorial Problems"[3]. A polynomial time solution to solving one, would be a solution to solve them all. And because all NP problems can be reduced to NP-Complete problems, the polynomial solution would solve all NP problems.

This algorithm I have presented, is one which runs in polynomial time no matter the input size. There is no branching, backtracking, guessing or heuristics involved and certainly no number $x^n$ calculations. This polynomial time algorithm solves k-clique, maximum clique and list of all maximal cliques, and therefore solves all NP problems. Therefore, P = NP.

Q.E.D.